\documentclass[%
 aip,
rsi,%
 amsmath,amssymb,
 reprint,%
]{revtex4-1}

\usepackage{graphicx}
\usepackage{dcolumn}
\usepackage{bm}
\usepackage{xcolor}

\graphicspath{{Figures/}}

\renewcommand{\d}{{\rm d}}
\newcommand{\e}{{\rm e}}
\newcommand{\FD}[2]{\frac{\d #1}{\d #2}}

\begin{document}

\preprint{AIP/123-QED}

\title[Mean-field approximation for networks with synchrony-driven adaptive coupling]{Mean-field approximation for networks with synchrony-driven adaptive coupling}
\author{N. Fennelly}
\thanks{joint first author}
 \affiliation{School of Mathematics and Statistics, University College Dublin, Ireland}
 
 \author{A. Neff}
 \thanks{joint first author}
 \affiliation{School of Mathematics, University of Edinburgh, UK}

 \author{R. Lambiotte}
 \affiliation{Mathematical Institute, University of Oxford, UK}
 
  \author{A. Keane}
 \affiliation{School of Mathematical Sciences, University College Cork, Ireland}
  
\author{\'{A}. Byrne}
 \email{aine.byrne@ucd.ie}
 \affiliation{School of Mathematics and Statistics, University College Dublin, Ireland}

\date{\today}

\begin{abstract}
Synaptic plasticity is a key component of neuronal dynamics, describing the process by which the connections between neurons change in response to experiences. In this study, we extend a network model of $\theta$-neuron oscillators to include a realistic form of adaptive plasticity. In place of the less tractable spike-timing-dependent plasticity, we employ recently validated phase-difference-dependent plasticity rules, which adjust coupling strengths based on the relative phases of $\theta$-neuron oscillators. We investigate two approaches for implementing this plasticity: pairwise coupling strength updates and global coupling strength updates. A mean-field approximation of the system is derived and we investigate its validity through comparison with the $\theta$-neuron simulations across various stability states. The synchrony of the system is examined using the Kuramoto order parameter. A bifurcation analysis, by means of numerical continuation and the calculation of maximal Lyapunov exponents, reveals interesting phenomena, including bistability and evidence of period-doubling and boundary crisis routes to chaos, that would otherwise not exist in the absence of adaptive coupling.

\end{abstract}

\maketitle

\begin{quotation}
A population of coupled neurons with synaptic plasticity may be modelled as an \emph{adaptive dynamical network} consisting of a very large number of excitable nodes and/or oscillators. One approach to circumvent the complexity of the full network is to consider the mean-field dynamics. Here, we introduce a mean-field model for a population of $\theta$-neurons that are adaptively coupled according to a phase-difference-dependent plasticity rule.  In this context, the mean-field dynamics describe important properties of neural synchrony and conductance. We analyse the validity of the model and discover a rich tapestry of dynamics, including multiple routes to chaos, induced by the adaptive coupling used to represent synaptic plasticity. 


\end{quotation}

\section{\label{sec:intro} Introduction}
Synaptic plasticity is the primary mechanism by which the brain learns, stores information, and adapts to new experiences. There is a plethora of experimental evidence confirming that the efficacy of synaptic connections are strengthened and weakened in response to experience and activity \cite{Bi1998, Markram1997, Feldman2000, Froemke2002, Cassenaer2007, Sgritta2017, Abbott2000, Mishra2016, Turrigiano2000}. 
The study of synaptic plasticity is crucial for deciphering the neural basis of cognitive functions and for developing therapeutic strategies for neurological disorders. Brain stimulation, both invasive and non-invasive, has been shown to induce lasting changes in the underlying neural circuit \cite{Kricheldorff2022}. Transcranial magnetic stimulation (TMS) is showing promising cognitive and behavioural responses in post-stroke rehabilitation\cite{Starosta2022} and patients with depression\cite{Fitzgerald2021}, but an understanding of how TMS evokes these responses and how best to apply it remains elusive.

Mathematical models of synaptic plasticity can be traced back to Donald Hebb who proposed that the repeated and persistent co-activation of two or more neurons leads to a strengthening of their synaptic connection \cite{Hebb1949}. Since Hebb's 1949 postulate, there have been an abundance of mathematical models to describe how synaptic connections are strengthened and weakened (see review by Taherkhani \emph{et al.} \cite{Taherkhani2020}). Of particular relevance for this study are adaptive network models, which are often used as proxies to understand synaptic plasticity in the brain \cite{berner2023adaptive}.
In general, network adaptivity enriches the palette of dynamical behaviours in networks of oscillators, with the possible emergence of multistability \cite{Aoki2009}, frequency clustering \cite{aoki2015self} as well as the  coexistence of incoherence and synchronisation \cite{skardal2014complex}.
Adaptive neuronal oscillator models typically employ phase-difference-dependent plasticity (PDDP) to update the coupling strengths based on how similar or dissimilar the neurons' oscillatory phases are. 
Duchet \emph{et al.} recently showed that, for a network of Kuramoto oscillators, these PDDP rules can be used synonymously with spike-timing-dependent plasticity rules \cite{Duchet2023}. The authors also showed that for a symmetric PDDP rule the Ott-Antonsen ansatz \cite{Ott2008} could be used to derive a low-dimensional mean-field description of the system. In this study, we extend the work of Duchet \emph{et al.} to the more biologically realistic $\theta$-neuron model with first order synaptic dynamics to incorporate adaptive coupling between the neurons. 

The $\theta$-neuron model is known to admit an exact mean-field reduction \cite{Luke2013} and when synaptic dynamics are included the reduced system takes the form of a neural mass model driven by a dynamical variable for the population synchrony \cite{Byrne2017, Coombes2019, Byrne2020}. Neural mass models, also known as firing rate models, are a class of mean-field models regularly employed to study coarse-grained neural activity. In general, these models are phenomenological in nature with no direct link to the underlying biology. However, this new form of neural mass model, derived from a population of $\theta$-neurons and dubbed the next generation neural mass model, provides an explicit link between the underlying spiking network and the measurable population-level activity. Similarly, Montbri\'{o} 
\emph{et al.} showed that the quadratic integrate-and-fire (QIF) model also admits an exact low dimensional description and derived a conformal mapping that links their mean-field model to the next generation neural mass model \cite{Montbrio2015}. 

As the $\theta$-neuron and QIF models include accurate descriptions of the synaptic interactions between cortical cells, it is possible to include plasticity, and other neural mechanisms, in a biologically plausible manner. Laing added a spatial dependence to the neurons' activity to derive an exact neural field model \cite{Laing2014}, and later extended this work to include and study the effect of gap junction coupling \cite{Laing2015}. 
Taher \emph{et al.} extended the model to include short-term synaptic facilitation and depression, and study working memory \cite{Taher2020}, while Gast and colleagues added short-term adaptation mechanisms to study bursting dynamics \cite{Gast2020, Gast2021a}. 

In the next section we introduce the mean-field model for a population of $\theta$-neurons and incorporate a PDDP rule to represent adaptive coupling strengths within the network. The resulting model is studied and compared to corresponding models of full network dynamics in Section~\ref{sec:comparison} in order to validate its accuracy. A bifurcation analysis of the mean-field model is presented in Section~\ref{sec:analysis}, followed by a more general discussion of the results in Section~\ref{sec:discuss}.

\section{\label{sec:model_deriv}Model derivation}
We begin by considering a network of $N$ synaptically coupled $\theta$-neurons. Their dynamics are described by
\begin{align}
\tau_m\FD{{\theta}_j}{t} &= (1-\cos\theta_j) + (1+\cos\theta_j)(\eta_j + s_j v_{\text{syn}}) \nonumber\\
&\phantom{.}\hspace{3cm}- s_j \sin\theta_j, \label{eq:theta_network} \\
\tau_s\FD{s_j}{t} & = - s_j + \frac{1}{N} \sum_{l=1}^N \sum_m k_{lj} \delta(t-t_l^m) ,
\label{eq:s_network}
\end{align}
where $\theta_j(t)$ is the phase and $s_j(t)$ is the synaptic conductance of neuron $j$. Parameters $\tau_m$ and $\tau_s$ are the membrane and synaptic timescales, respectively, $\eta_j$ is the background drive, $v_{\text{syn}}$ is the synaptic reversal potential and $k_{jl}$ is the strength of the synaptic connection from neuron $j$ to neuron $l$.
The $\theta$-neuron model is formally equivalent to the quadratic integrate-and-fire (QIF) model, under the transformation $v=\tan\frac{\theta}{2}$. We include conductance based synaptic coupling to arrive at \eqref{eq:theta_network} (see Appendix \ref{app:QIF} for the equivalent QIF network). 

We employ the PDDP rule of Seliger \emph{et al.} \cite{Seliger2002} to update the synaptic coupling strengths based on the phase differences between pairs of neurons
\begin{align}
\FD{k_{jl}}{t} & = \epsilon\left(-k_{jl}  + \alpha\cos(\theta_l-\theta_j)\right),
\label{eq:k_network}
\end{align}
where $\alpha$ controls the strength of plasticity relative to the intrinsic decay of synaptic strength and $\epsilon$ is the plasticity timescale.

For a network of $\theta$-neurons with uniform coupling strengths, $k_{jl}=k, \forall \{j,l\}$, the mean-field dynamics are given as
\begin{align}
\tau_m\FD{z}{t} &= -i\frac{(z-1)^2}{2}+\frac{(z+1)^2}{2}\left[-\Delta + i \eta_0 + i s v_{\text{syn}} \right] \nonumber \\
&\phantom{.}\hspace{4cm}- \frac{z^2-1}{2} s, \label{eq:z_MF} \\
\tau_s\FD{s}{t} & = - s + \frac{k}{\pi\tau_m} \frac{1-\left|z\right|^2}{1+z+\overline{z}+\left|z\right|^2}.\label{eq:s_MF}
\end{align}
where $s(t)=\frac{1}{N}\sum_{j=1}^N s_j(t)$ is the mean synaptic conductance and $z(t)=\frac{1}{N}\sum_{j=1}^N e^{i\theta_j(t)}= R(t) e^{i\phi(t)}$ is the Kuramoto order parameter.
The Kuramoto order parameter encapsulates the average phase of the neurons, $\phi(t)$, and the phase coherence, $R(t)$, i.e. the level of synchrony. The parameters $\eta_0$ and $\Delta$ describe the mean and width of the Lorentzian distribution, respectively, of the background drives.

Equations~\eqref{eq:z_MF}--\eqref{eq:s_MF} is the mean-field model derived by Coombes and Byrne \cite{Coombes2019} in the case of a fixed network topology, with first order synaptic coupling rather than second. 
To derive a mean-field approximation for the network with PDDP, we must construct an equation for the evolution of the mean coupling strength 
$\hat{k} = \frac{1}{N^2}\sum_{j=1}^N\sum_{l=1}^N k_{jl}$,
\begin{align}
\FD{\hat{k} }{t}& = \epsilon\left(-\hat{k}  +  \frac{\alpha }{N^2}\sum_{j=1}^N \sum_{l=1}^N \cos(\theta_l-\theta_j)\right) \nonumber\\
& = \epsilon\left(-\hat{k}  +  \alpha\left(\frac{1}{N}\sum_{j=1}^N e^{i\theta_j}\right) \left(\frac{1}{N}\sum_{l=1}^N e^{-i\theta_l}\right) \right) \nonumber\\
& = \epsilon\left(-\hat{k}  + \alpha |z|^2\right).
\label{eq:k_average}
\end{align}
Notice how the  mean coupling strength is updated based on the magnitude of the Kuramoto order parameter, i.e. the level of synchrony. Hence, the more synchronised the neural activity is, the stronger the coupling between these neurons becomes. This aligns with the Hebbian theory that \emph{neurons that fire together, wire together}. In particular, it accounts for the higher order coupling updates, increasing the strength of all connections when the network is highly synchronised rather than a single pair.
By making this approximation, we assume that the synaptic coupling strengths follow a unimodal distribution and, as such, forgo the possibility of studying clustering behaviour.

The mean-field dynamics are given by \eqref{eq:z_MF}--\eqref{eq:k_average} with $k=\hat{k}$. We note that \eqref{eq:k_average} was derived previously by Duchet \emph{et al.}, but coupled with another model for the synchronisation of  oscillators, i.e. the Kuramoto model. Here, we combine it with the more biologically plausible $\theta$-neuron network with first order synaptic coupling. The increased complexity of the oscillator model results in a rich bifurcation structure and a diverse set of behaviours (see Section \ref{sec:analysis}).

\begin{center}
\begin{figure*}
\includegraphics[width=1\linewidth]{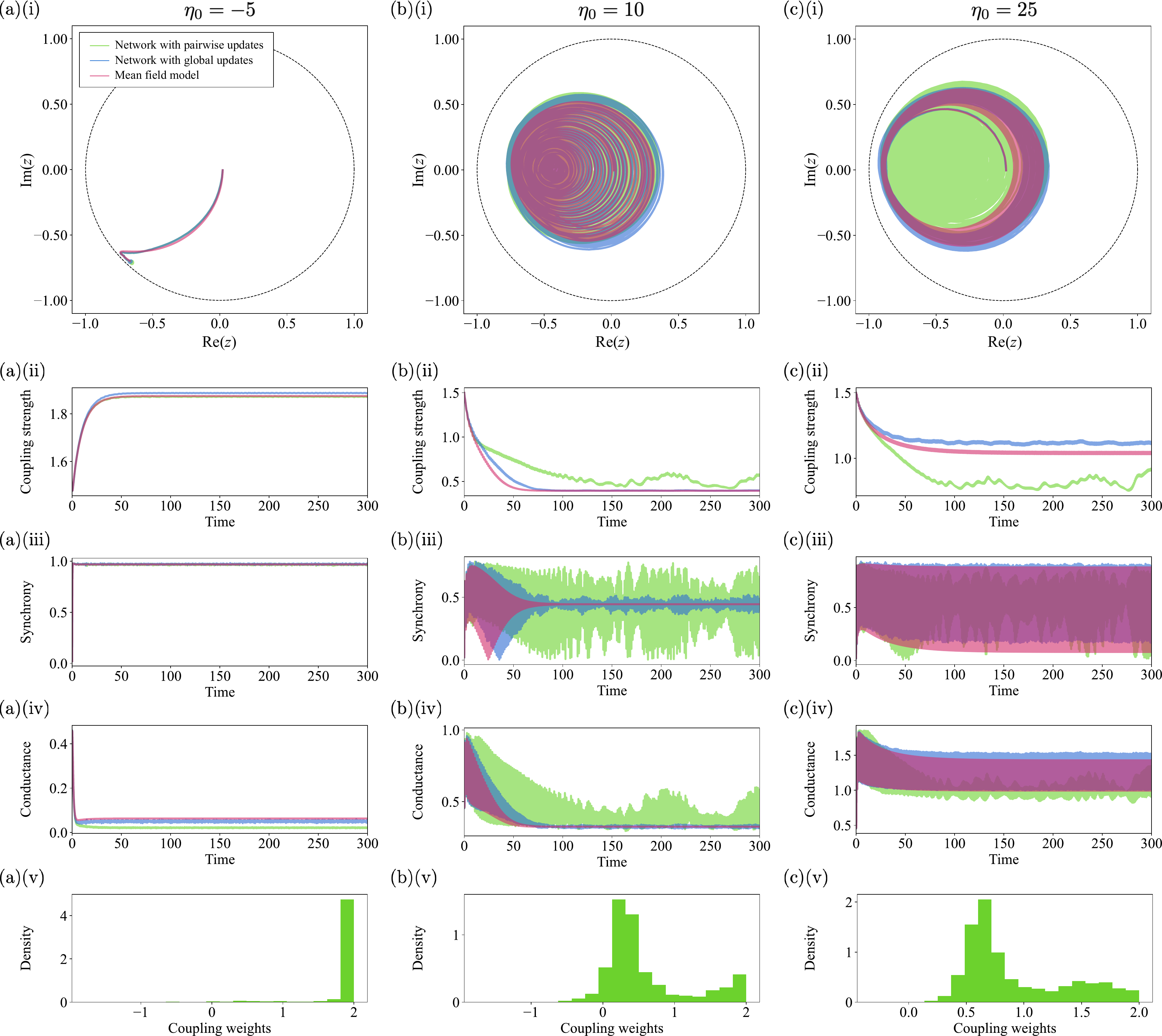}
\caption{Comparison of mean field model with networks of 1000 $\theta$-neurons in 
different regimes - (a) stable node ($\eta_0=-5$), (b) stable spiral ($\eta_0=10$), (c) limit cycle ($\eta_0=25$). Network simulations with pairwise (global) updates are shown in green (blue) and the mean-field dynamics are shown in pink. Other parameter values: $\Delta=0.5$, $v_{\text{syn}}=-10$, $\tau_m=1$, $\tau_s=1$, $\alpha=2$, $\epsilon=0.1$.}\label{fig:mean_field_validation}
\end{figure*}
\end{center}

\section{\label{sec:comparison}Mean-field validation}
We study the $\theta$-neuron network model \eqref{eq:theta_network}--\eqref{eq:s_network}, where the pairwise coupling strengths are updated according to \eqref{eq:k_network} (i.e. network with pairwise or `local' updates). We compare the results with the same $\theta$-neuron network, where all coupling strengths are now updated based on the mean evolution equation \eqref{eq:k_average} (i.e. network with global updates), as well as the mean-field system \eqref{eq:z_MF}--\eqref{eq:k_average} with $k=\hat{k}$.
Figure~\ref{fig:mean_field_validation} displays results from numerical simulations using a network size of 1000. Due to the presence of discontinuities as neurons fire in the full network models, we use Euler's method with a sufficiently small step size. Each column corresponds to a different value of $\eta_0$ in order to capture regimes of qualitatively different dynamical behaviour. The green, blue and pink curves show the evolution of the dynamic variables of the network with pairwise updates, global updates and of the mean-field model, respectively. The bottom row shows the coupling weight distributions of the network with pairwise updates.

For fixed point dynamics, the mean-field model is a good approximation for both the network with local updates and the network with global updates, especially when the fixed point lies close to the boundary $|z|=1$ (Fig.~\ref{fig:mean_field_validation}(a)). 
When $|z|=1$, the neurons are perfectly synchronised and the phase differences are zero. While for $|z|\approx1$, there is very little heterogeneity in the phase differences and the distribution of coupling weights has a single narrow peak at the mean coupling strength of $\sim1.9$ (Fig. \ref{fig:mean_field_validation}(a)(v)). Hence, the global coupling strength update rule is a close approximation for the local pairwise coupling strength update rule in this regime. 
Although the population-level dynamics tend to a fixed point, the neurons are not necessarily quiescent.
As shown by Byrne \emph{et al.} \cite{Byrne2017}, the firing rate is positive for all $|z|<1$ and increases as the dynamics move towards $z=\e^{i\pi}$, where firing is maximal as all of the neurons cross threshold at exactly the same time (see Fig. 8 of Byrne \emph{et al.}\cite{Byrne2017}). In this case where $|z|\approx1$ and $\phi\neq\pi$ the majority of neurons are synchronised at rest, and as such, this fixed point corresponds to a population of predominantly quiescent neurons.

As the fixed point moves away from the $|z|=1$ boundary, the firing rate increases and the population of neurons switches to an active state. Moving away from the $|z|=1$ boundary also means a reduction in the level of synchrony, and as such,
the phases become more heterogeneous and the dynamics of the networks with local and global updates begin to diverge (Fig.~\ref{fig:mean_field_validation}(b)). However, networks with global updates still closely match the mean-field dynamics. The distribution of coupling weights starts to become bimodal (Fig. \ref{fig:mean_field_validation}(b)(v)), suggesting that the oscillators may be forming clusters. Although the mean coupling strength is higher in the network with local updates (Fig. \ref{fig:mean_field_validation}(b)(ii)), the Kuramoto order parameter still fluctuates around the mean-field fixed point value. This is particularly apparent in the synchrony time course (Fig. \ref{fig:mean_field_validation}(b)(iii)). 

For limit cycle dynamics, the mean-field model is a good approximation for the network with global updates only (Fig.~\ref{fig:mean_field_validation}(c)). Although the trajectories do not match for the network with local updates, they lie in roughly the same region of the Kuramoto order parameter phase plane (Fig. \ref{fig:mean_field_validation}(c)(i)). 
The distribution of coupling weights is now clearly bimodal, with a peak at roughly  0.7 and 1.6 (Fig. \ref{fig:mean_field_validation}(c)(v)). The mean-field model slightly underestimates the mean coupling strength of the network with global updates and the oscillatory amplitude of the conductance, and slightly overestimates the oscillatory amplitude of the level of synchrony. However, the dynamics are comparable and qualitatively similar. We note that increasing the number of neurons and reducing the step size for the network simulation results in a closer fit to the mean-field model.

\begin{center}
\begin{figure}[h!]
\includegraphics[width=0.85\linewidth]{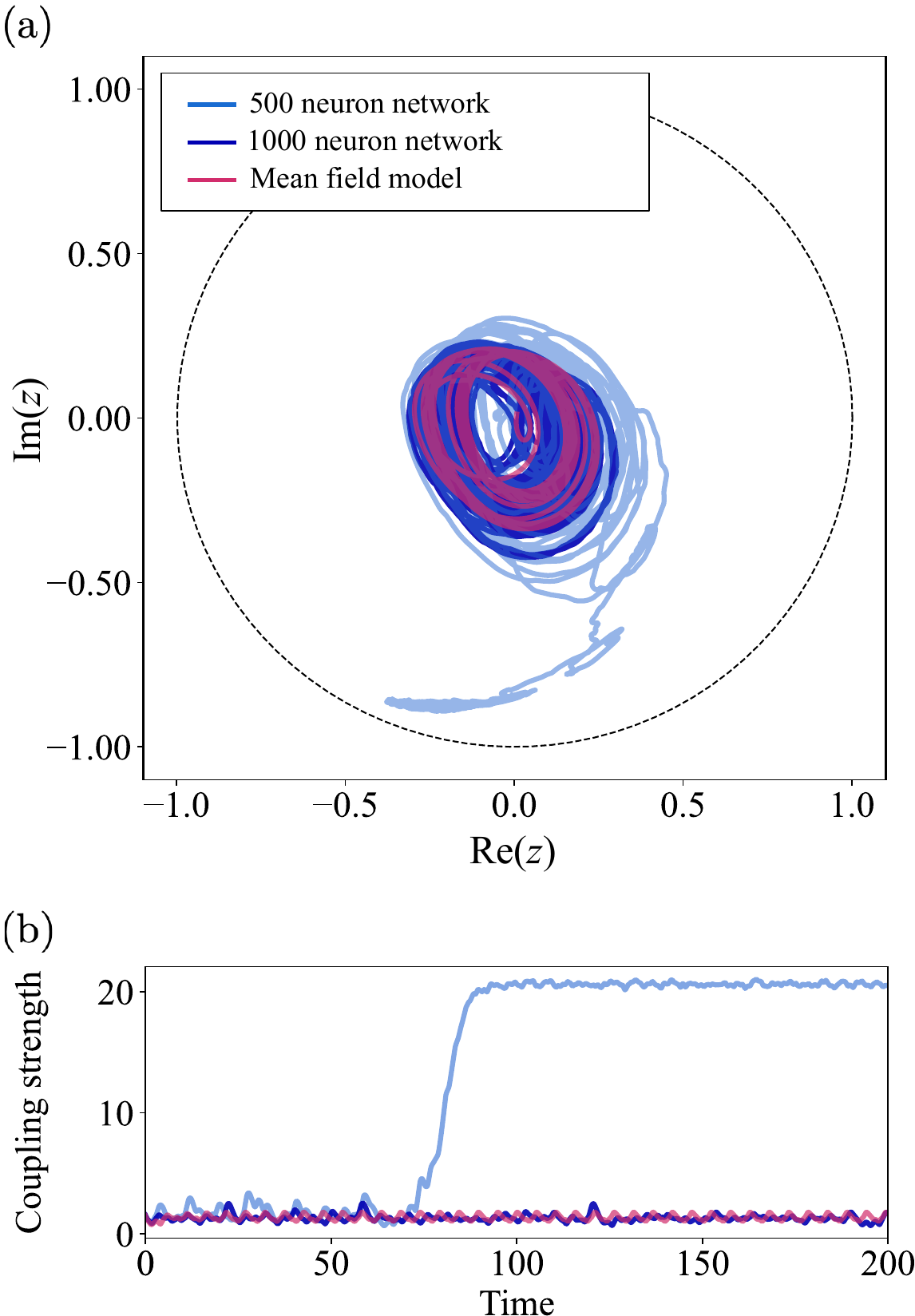}
\caption{Bistability. Finite noise fluctuations in network with global updates drives system to alternative stable attractor. Network simulations with 500 (1000) neurons shown in light (dark) blue and mean-field simulation shown in pink. Parameter values: $\eta=5.25$, $\Delta=0.5$, $v_{\text{syn}}=-10$, $\tau_m=1$, $\tau_s=1$, $\alpha=25$, $\epsilon=0.5$}
\label{fig:bistability}
\end{figure}
\end{center}

Interestingly, when comparing the dynamics of the mean-field system to the network with global updates, we observed transitions in the network simulations away from the mean-field dynamics (Fig. \ref{fig:bistability}). This indicated the existence of another stable attractor, which was confirmed via bifurcation analysis (see Section \ref{sec:analysis}). The behaviour in Fig.~\ref{fig:bistability} suggests that the finite size fluctuations in the network model may push the system out of the basin of attraction of the limit cycle attractor and into the basin of attraction of the stable fixed point where the network is more strongly coupled (light blue curve). Increasing the network size from 500 to 1000 reduces the finite size fluctuations, and, as such, the network dynamics remain within the basin of attraction of the limit cycle attractor (dark blue curve). 
In order to gain insights into the potential for multistability and complex behaviour, we conduct a bifurcation analysis in the following section.

\section{\label{sec:analysis}Bifurcation analysis}

We perform a bifurcation analysis of the mean-field approximation of the network with PDDP \eqref{eq:z_MF}--\eqref{eq:k_average} by means of numerical simulation and continuation. We use the 4\textsuperscript{th} order Runge-Kutta  method to find all stable behaviour, including fixed point, periodic, quasi-periodic and chaotic attractors, and generally discard transient behaviour, unless stated otherwise. We use the Matlab-based continuation package MatCont \cite{dhooge2008new} to obtain unstable fixed point solutions and periodic orbits, as well as to calculate their stability properties and identify bifurcations. This preliminary analysis has the specific aim to highlight the richness of dynamics that emerges as the coupling is allowed to become \emph{adaptive}. Avenues for a more comprehensive analysis are discussed in the following section. Our analysis focuses on the mean background drive $\eta_0$, which is a bifurcation parameter commonly used in previous studies of the model without PDDP.

\begin{figure}[t]
\includegraphics[width=1\columnwidth]{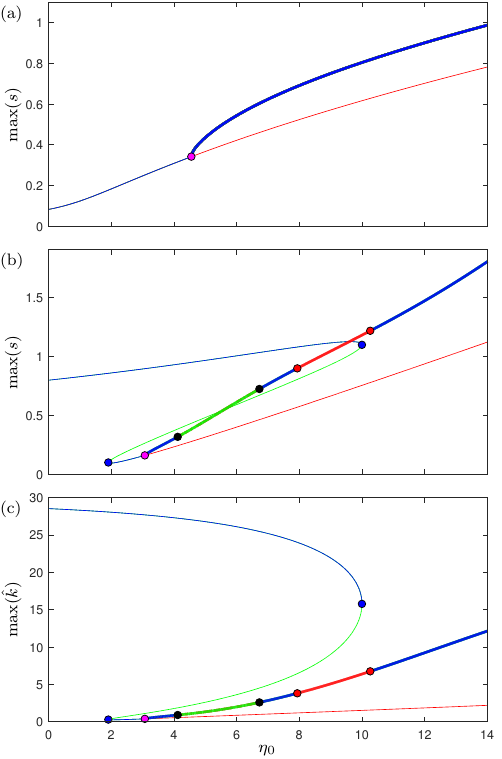}
\caption{Bifurcation diagram in $\eta_0$ for fixed coupling ($k=1$, $\epsilon=0$) in panel~(a) and adaptive coupling ($\alpha=30$, $\epsilon=0.5$) in panels~(b)--(c). Thin curves represent fixed points with zero (blue), one (green) or two (red) unstable eigendirections. Blue and magenta circles represent saddle-node and Hopf bifurcations, respectively. Thick curves represent periodic orbits with zero (blue), one (green) or two (red) unstable Floquet multipliers. Black and red circles represent period-doubling and torus bifurcations, respectively. 
Other parameter values as in Fig.~\ref{fig:bistability}.} 
\label{fig:bifs_eta_alpha30}
\end{figure}

To begin, we present a bifurcation diagram of the model without PDDP (i.e. $\epsilon=0$ and fixed $k=1$) in order to demonstrate the typical behaviour of the model with \emph{fixed} coupling. Figure~\ref{fig:bifs_eta_alpha30}(a) shows a thin curve representing fixed points, in terms of synaptic conductance $s$, that are either stable (blue) with zero unstable eigendirections or saddle (red) with two unstable eigendirections. 
The thick blue curve shows the maximum $s$ value of stable periodic orbits emerging from the Hopf bifurcation, represented by the magenta circle. This bifurcation diagram is in agreement with previous studies that have shown that the system may undergo a supercritical Hopf bifurcation for appropriate parameter values 
\cite{Luke2013, Coombes2019}. This relatively simple behaviour contrasts the complex dynamics that emerge when adaptive plasticity is incorporated into the model.

Figure~\ref{fig:bifs_eta_alpha30}(b) shows the bifurcation diagram for the same mean-field model, now with adaptive coupling ($\alpha=30$ and $\epsilon=0.5$). The thin green curve represents fixed points with one unstable eigendirection. The thick green and red curves represent saddle periodic orbits with one and two unstable Floquet multipliers, respectively. Blue, black, and red circles indicate saddle-node, period-doubling and torus bifurcations. Panel~(c) shows the same bifurcation diagram in terms of mean coupling strength $\hat{k}$. 

For $\eta_0 = 0$, when the mean background drive is zero, the only attractor is a fixed point that represents a strongly coupled network. As $\eta_0$ is increased, the coupling strength decreases until the branch of equilibria loses stability at a saddle-node bifurcation at $\eta_0 \approx 10$. A branch of saddle fixed points leads to stable fixed points, representing a weakly coupled network, via another saddle-node bifurcation. 
The lower branch of stable fixed points is short-lived, since a supercritical Hopf bifurcation is encountered, producing a branch of stable periodic orbits. The periodic orbits become unstable twice as $\eta_0$ increases: they gain one unstable Floquet multiplier between a pair of period-doubling bifurcations and two unstable Floquet multipliers between a pair of torus bifurcations.

\begin{figure}[t]
\includegraphics[width=1\columnwidth]{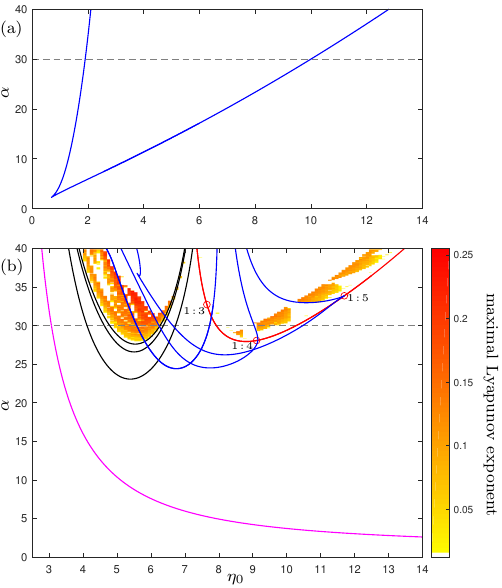}
\caption{Bifurcation set in the $(\eta_0,\alpha)$-plane involving fixed points only (a) and periodic orbits (b). Panel~(a) shows blue curves of saddle-node bifurcations of fixed points. Panel~(b) shows curves of Hopf bifurcations (magenta), period-doubling bifurcations (black), torus bifurcations (red) and saddle-node bifurcations of periodic orbits (blue). The labels on the torus bifurcation curve show the root points of the corresponding ${p\!:\!q}$ Arnold tongues. The shaded regions indicate positive maximal Lyapunov exponents. The dashed grey line indicates the value of $\alpha$ used for Figs.~\ref{fig:bifs_eta_alpha30} and~\ref{fig:PDcascade}.
Other parameter values as in Fig.~\ref{fig:bistability}.}
\label{fig:bifs_eta_alpha}
\end{figure}

The numerous bifurcations evident in Fig.~\ref{fig:bifs_eta_alpha30}(b)--(c) spurred an investigation of the bifurcation set in the ($\eta_0$, $\alpha$)-plane. We identified $\alpha$ as a suitable second bifurcation parameter, since it relates to the effect of the adaptive coupling, as it represents an upper bound on the coupling strength. 
Figure~\ref{fig:bifs_eta_alpha}(a) shows the saddle-node bifurcations of fixed points, observed in Fig.~\ref{fig:bifs_eta_alpha30}(b)--(c), that meet at a cusp bifurcation. Inside the two curves exist a region of bistability. For the parameter values considered here, the upper branch of fixed points remains stable, while the lower branch undergoes numerous bifurcations. 
Figure~\ref{fig:bifs_eta_alpha}(b) shows the bifurcations that emerge from the lower branch in Fig.~\ref{fig:bifs_eta_alpha30}(b)--(c) continued in the ($\eta_0$, $\alpha$)-plane. In other words, the Hopf (magenta), period doubling (black) and torus (red) bifurcation points in Fig.~\ref{fig:bifs_eta_alpha30}(b)--(c) now correspond to curves in Fig.~\ref{fig:bifs_eta_alpha}(b). 
The dashed grey line indicates the value of $\alpha$ used in Fig.~\ref{fig:bifs_eta_alpha30}(c).
In addition, we have blue curves representing saddle-node bifurcations of periodic orbits, additional period doubling curves in black, and shaded regions of the plane displaying maximal Lyapunov exponents. For clarity, we only show the positive maximal Lyapunov exponents.

Figure~\ref{fig:bifs_eta_alpha}(b) provides valuable insights into the behaviour of the system. 
Firstly, we see that the pair of period doubling bifurcations shown in Fig.~\ref{fig:bifs_eta_alpha30}(c) belong to the same bifurcation curve in Fig.~\ref{fig:bifs_eta_alpha}(b). We show further period doubling bifurcation curves to demonstrate the period doubling cascade that gives rise to a region of chaotic behaviour. We find that solutions within this region possess positive maximal Lyapunov exponents, providing evidence that these solutions are indeed chaotic. Note that the maximal Lyapunov exponents are calculated by incrementally increasing $\alpha$ for fixed values of $\eta_0$, so it is plausible that some positive maximal Lyapunov exponents are overlooked due to multistability.

\begin{figure}[t]
\includegraphics[width=1\columnwidth]{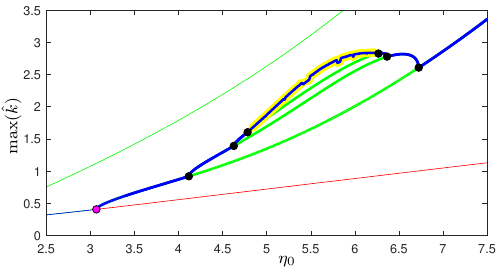}
\caption{Period doubling cascade in $\eta_0$ for $\alpha=30$ in terms of max($\hat{k}$). Thin curves represent fixed points with zero (blue), one (green) or two (red) unstable eigendirections, whereby the magenta circle indicate a Hopf bifurcation. Thick curves represent periodic orbits with zero (blue) or one (green) unstable eigendirections, whereby black circles indicate period-doubling bifurcations. Chaotic solutions are shaded yellow.
Other parameter values as in Fig.~\ref{fig:bistability}.} 
\label{fig:PDcascade}
\end{figure}

A more detailed bifurcation diagram in $\eta_0$ for fixed $\alpha=30$, which includes the period-doubling route to chaos, is shown in Fig.~\ref{fig:PDcascade}. The solutions with a positive maximal Lyapunov exponent are highlighted yellow, revealing the branch of chaotic attractors between the sets of period doubling cascades.

Returning to Fig.~\ref{fig:bifs_eta_alpha}(b), as $\eta_0$ is increased further, the periodic orbit born at the Hopf bifurcation loses stability at a torus bifurcation. 
Our simulations reveal that small invariant tori emerge beyond the torus bifurcation curve, where the periodic orbit is unstable. These tori persist in long simulation runs, suggesting that the torus bifurcations are supercritical (for the range of parameter values considered here). Aside from quasiperiodic solutions on the invariant tori, they will also have associated ${p\!:\!q}$ Arnold tongues, as labelled in Fig.~\ref{fig:bifs_eta_alpha}(b).
Generally, where the underlying torus is stable, the Arnold tongue will contain a pair of stable and saddle ${p\!:\!q}$ periodic orbits. Boundaries of the tongue are given by saddle-node bifurcations where these two periodic orbits meet and disappear. Here, we show three example Arnold tongues. In accordance with theory \cite{arnold2012geometrical,kuznetsov1998elements}, the ${1\!:\!5}$ Arnold tongue is rooted at the point on the torus bifurcation curve where its rotation number (or winding number) becomes $1/5$, indicated by the red circle. In theory, there exist an infinite number of Arnold tongues, each rooted at an associated point along the torus bifurcation curve with a rational rotation number. The ${1\!:\!3}$ and ${1\!:\!4}$ Arnold tongues, on the other hand, do not appear to be rooted at their associated resonance points along the torus bifurcation curve. This suggests that these Arnold tongues possess a more complicated geometry, as discussed in the following section.

\begin{figure}[t]
\includegraphics[width=1\columnwidth]{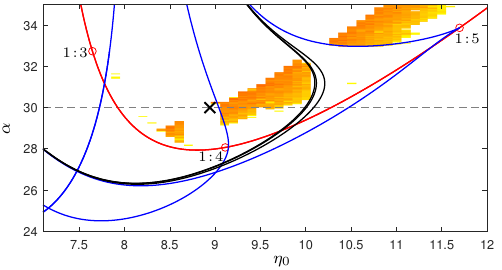}
\caption{Zoom-in on the ${1\!:\!5}$ Arnold tongue in Fig.~\ref{fig:bifs_eta_alpha} with (black) period doubling bifurcation curves. The black cross indicates the parameter values used in Fig.~\ref{fig:transients}.
Other parameter values as in Fig.~\ref{fig:bistability}.}
\label{fig:bifs_eta_alpha_zoom}
\end{figure}

\begin{figure}[t]
\includegraphics[width=1\columnwidth]{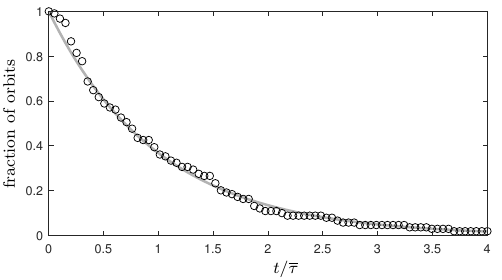}
\caption{
Fraction of transients remaining on ghost of the chaotic attractor as a function of time $t$ from numerical simulation (black circles) and theory (grey curve). $5000$ transients are calculated for $\eta_0=8.94$ and $\alpha=30$ using random initial conditions within the former basin of attraction of the chaotic attractor. The average time spent on the ghost of the chaotic attractor is $\overline{\tau}=2140$. 
Other parameter values as in Fig.~\ref{fig:bistability}.
}
\label{fig:transients}
\end{figure}

The maximal Lyapunov exponents in Fig.~\ref{fig:bifs_eta_alpha}(b) also indicate regions of chaotic behaviour due to dynamics on tori for larger values of $\eta_0$. The emergence of chaotic regimes, resulting from the overlapping of Arnold tongues, has been observed in many systems throughout the literature; for example, \cite{cumming1987deviations,keane2016investigating,heltberg2021tale}. The chaotic attractor may appear and disappear via multiple routes. For example, in Fig.~\ref{fig:bifs_eta_alpha_zoom} we show three period doubling curves that exist within the ${1\!:\!5}$ Arnold tongue, representing a cascade of period doubling bifurcations that leads to chaotic behaviour. In some cases the disappearance of chaotic attractors appear to coincide with saddle-node bifurcations of periodic orbits. In such cases, our simulations reveal very long chaotic transients. This suggests that chaotic attractors may disappear at boundary crises, where a chaotic attractor meets the stable manifold of a saddle invariant set \cite{grebogi1983crises}. To provide further evidence of this, Fig.~\ref{fig:transients} displays the results of 5000 simulations for parameter values near where the chaotic attractor disappears ($\eta_0=8.94$ and $\alpha=30$; black cross in Fig.~\ref{fig:bifs_eta_alpha_zoom}) using randomly chosen initial conditions. They are chosen from the former basin of attraction of the chaotic attractor. The black circles on the plot show the fraction of the 5000 simulations that remain on the ghost of the chaotic attractor as a function of time. The horizontal axis is scaled by the average chaotic transient time $\overline{\tau}=2140$. We see that all trajectories eventually leave the ghost of the chaotic attractor. They all settle to the stable ${1\!:\!5}$ periodic orbit. The black circles are in good agreement with the grey curve, which indicates the theoretical (exponential) probability distribution of chaotic transient lengths in the nearby vicinity of a boundary crisis \cite{grebogi1983crises}.

\section{\label{sec:discuss} Discussion}
In this work, we derive a low-dimensional description for a network of $\theta$-neurons with first order synapses and adaptive coupling. We compare the dynamics of the reduced system to the $\theta$-neuron network with both local pairwise coupling updates and with global population-wide coupling updates, and find that the reduced system accurately describes the evolution of the synchrony and the mean coupling strength for the global updates. For the local update rule, the reduced system only reproduces the network dynamics when the neurons do not cluster, and therefore the coupling weights form a unimodal distribution.
Due to the $k=\hat{k}$ assumption, the accuracy of the mean-field model is also related to how narrow this distribution is.

We presented a preliminary bifurcation analysis of the mean-field model in order to highlight the relatively complicated dynamics that appears once the coupling is modelled to be adaptive. 
It is evident in Fig.~\ref{fig:bifs_eta_alpha}(b) that the relatively interesting behaviour, compared to the fixed coupling case, only occurs when $\alpha$ is sufficiently large. For smaller values of $\alpha$, only a Hopf bifurcation is encountered as $\eta_0$ is varied, such that the behaviour is similar to the fixed coupling case.
For larger values of $\alpha$, the periodic orbit created at the Hopf bifurcation undergoes further bifurcations, leading to two distinct regimes of chaotic behaviour. One is due to a period doubling cascade. The other is due to overlapping Arnold tongues created along the torus bifurcation curve. 

Phase-difference-dependent plasticity (PDDP) rules are commonly used in adaptive network models as a proxy for spike-time-dependent plasticity (STDP) \cite{Seliger2002, Aoki2009, Lucken2016, Berner2019}. The validity of this approximation was studied by Duchet \emph{et al.}, who constructed PDDP rules to fit both symmetric and asymmetric STDP rules, and showed that the emergent dynamics and coupling weights were closely matched \cite{Duchet2023}. In making the mean-field reduction, we lose any notion of spike times and are forced to rely on synchrony as a measure of how similar or dissimilar the activities are. For single-cluster states, where the activities of all neurons are similar, the reduced model still provides an accurate description of the network with STDP-like plasticity. However, if the plasticity induces clustering, population-level synchrony is no longer an accurate surrogate for spike-time difference or phase difference. The framework presented here is readily extendable to study multi-cluster states via the introduction of multiple populations and/or higher-order Kuramoto–Daido order parameters.

Although STDP has dominated the synaptic plasticity literature, particularly in the field of mathematical and computational neuroscience, it is not the only mechanism for activity dependent changes in synaptic coupling strengths \cite{Citri2008}. Synaptic scaling is a global plasticity mechanism that strengthens or weakens all synapses by the same factor. This homeostatic plasticity mechanism acts to stabilise and regulate the network activity, to ensure normal propagation of signals through the network \cite{Turrigiano2012,Berberich2017}. Our global update rule takes the form of a synaptic scaling law which regulates the network activity based on the population synchrony. In its current form, the plasticity rule promotes synchrony, increasing the coupling strength when the neurons are synchronised, but one could easily add a phase shift into the original PDDP rule to modify it to promote asynchrony or other partially synchronised states.

We note that the PDDP rule considered here pays no recourse to the actual firing time and instead updates the coupling strengths based on the phase difference at all values of time. Duchet \emph{et al.} showed that this simple version of the PDDP rule provided a reasonable approximation of the network and coupling weight dynamics for a network of Kuramoto oscillators with STDP \cite{Duchet2023}. The authors also introduced an event-based PDDP rule whereby the coupling weights are updated according to phase differences but only the phase corresponding to spiking. Given that Duchet \emph{et al.} did not find distinctly different dynamics for the event-based rule, we chose to use the simple PDDP in this work, but note that the work here could easily be extended to include the event-based rule.

There are numerous intriguing avenues for future work into the analysis of the mean-field model considered in this paper. It is still not clear how the stable ${1\!:\!3}$ and ${1\!:\!4}$ Arnold tongues presented in Fig.~\ref{fig:bifs_eta_alpha}(b) emerge from the torus bifurcation curve. We do detect curves of torus bifurcations and neutral saddles within these Arnold tongues that could belong to so-called Chenciner bubbles, suggesting that the Arnold tongues attach to the torus bifurcation curve via folds of tori that connect regimes of stable and saddle tori. To show this clearly would require a careful and intricate numerical analysis, as demonstrated by Keane and Krauskopf\cite{keane2018chenciner}.

\begin{figure}[t]
\includegraphics[width=1\columnwidth]{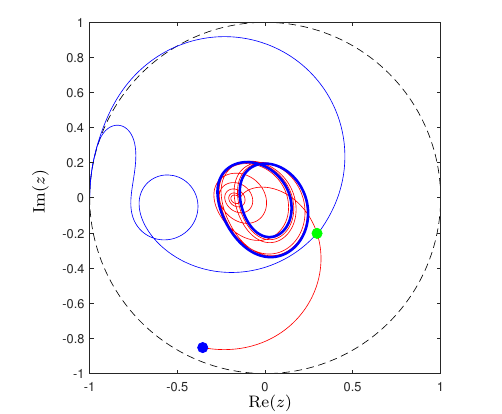}
\caption{$z$-plane of the mean-field model for parameter values used in Fig.~\ref{fig:bistability}. Blue/green circles represent stable/saddle fixed points, while thick blue curve represents a periodic orbit. The thin red and blue curves are approximations of the unstable and strong stable manifolds of the saddle, respectively.}
\label{fig:zplane_eta525_alpha25}
\end{figure}

In Section \ref{sec:analysis} we provide evidence of a boundary crisis route to chaos (Fig.~\ref{fig:transients}). To clearly visualise such crises in this system would require the calculation of higher-dimensional saddle manifolds using advanced numerical techniques; for example, the work of Hasan \emph{et al.}\cite{hasan2018saddle} and Dankowicz \emph{et al.} \cite{dankowicz2020multidimensional}. A deeper appreciation of the role of saddle manifolds in this system could also be beneficial for understanding the limitations of the mean-field model. In particular, if an attractor of the mean-field model is near the boundary of its basin of attraction, then this mean-field solution may not correspond to a solution in the full network system, since the variability around its mean state may force the system into a different basin of attraction. For example, in Fig.~\ref{fig:bistability} we observe that the simulation of the mean-field model follows a periodic solution, while the full network simulation appears to approximately follow the periodic solution for some time before jumping to a different attractor with a larger coupling strength. Figure~\ref{fig:zplane_eta525_alpha25} shows the same phase space projection as Fig.~\ref{fig:bistability}(a), containing the relevant invariant sets of the mean-field model. The thick blue curve represents the stable periodic orbit, the blue/green circle represents the stable/saddle fixed point. The Jacobian of the model evaluated at the saddle has the following eigenvalues: $(-3.016, -0.5240\pm2.224i, 0.5915)$. As such, the saddle has a one-dimensional unstable manifold and a three-dimensional stable manifold. In Fig.~\ref{fig:zplane_eta525_alpha25} the thin red curve represents the unstable manifold of the saddle fixed point, found by numerically integrating the model forwards in time from the saddle. The thin blue curve is an approximation of the strong stable manifold, found by numerically integrating the model backwards in time from the saddle. Since the boundary between the basins of attraction is given by the stable manifold of the saddle, Fig.~\ref{fig:zplane_eta525_alpha25} paints a convincing picture that the jump observed in Fig.~\ref{fig:bistability} is due to the natural variability of the full-network model pushing the trajectory from the periodic orbit beyond the stable manifold of the saddle and, therefore, into the basin of attraction of the stable fixed point. However, the picture in Fig.~\ref{fig:zplane_eta525_alpha25} is somewhat misleading since the stable manifold of the saddle is three-dimensional and it only shows the one-dimensional strong stable manifold. Also, it cannot be ruled out that the close vicinity of the periodic orbit to the strong stable manifold of the saddle in Fig.~\ref{fig:zplane_eta525_alpha25} is due to a fortunate projection from four dimensions onto two dimensions. Hence, providing a clear and precise picture of how trajectories interact with saddle manifolds would require their calculation in higher dimensions using more advanced techniques.

Finally, it could be worthwhile to consider smaller values of $\epsilon$, representing a slower adaptive coupling rate. For the case of sufficiently small $\epsilon$, it would be useful to consider a multiple time-scale approach using geometric singular perturbation theory \cite{fenichel1979geometric, Jones1995}.

\begin{acknowledgments}
We thank Cris Hasan for helpful discussions about saddle manifolds in $\mathbb{R}^4$.
RL acknowledges support from the EPSRC grants EP/V013068/1, EP/V03474X/1 and EP/Y028872/1.
\end{acknowledgments}

\section*{References}
\bibliographystyle{unsrturl}

\begin{thebibliography}{10}

\bibitem{Bi1998}
Guo~Qiang Bi and Mu~Ming Poo.
\newblock {Synaptic modifications in cultured hippocampal neurons: Dependence
  on spike timing, synaptic strength, and postsynaptic cell type}.
\newblock {\em Journal of Neuroscience}, 18(24):10464--10472, 1998.
\newblock \href {https://doi.org/10.1523/jneurosci.18-24-10464.1998}
  {\path{doi:10.1523/jneurosci.18-24-10464.1998}}.

\bibitem{Markram1997}
Henry Markram, Joachim L{\"{u}}bke, Michael Frotscher, and Bert Sakmann.
\newblock {Regulation of synaptic efficacy by coincidence of postsynaptic APs
  and EPSPs}.
\newblock {\em Science}, 275(5297):213--215, 1997.
\newblock \href {https://doi.org/10.1126/science.275.5297.213}
  {\path{doi:10.1126/science.275.5297.213}}.

\bibitem{Feldman2000}
Daniel~E Feldman.
\newblock {Timing-based LTP and LTD at vertical inputs to layer II/III
  pyramidal cells in rat barrel cortex}.
\newblock {\em Neuron}, 27(1):45--56, 2000.
\newblock \href {https://doi.org/10.1016/S0896-6273(00)00008-8}
  {\path{doi:10.1016/S0896-6273(00)00008-8}}.

\bibitem{Froemke2002}
Robert~C. Froemke and Yang Dan.
\newblock {Spike-timing-dependent synaptic modification induced by natural
  spike trains}.
\newblock {\em Nature}, 416(6879):433--438, mar 2002.
\newblock \href {https://doi.org/10.1038/416433a} {\path{doi:10.1038/416433a}}.

\bibitem{Cassenaer2007}
Stijn Cassenaer and Gilles Laurent.
\newblock {Hebbian STDP in mushroom bodies facilitates the synchronous flow of
  olfactory information in locusts}.
\newblock {\em Nature}, 448(7154):709--713, jun 2007.
\newblock \href {https://doi.org/10.1038/nature05973}
  {\path{doi:10.1038/nature05973}}.

\bibitem{Sgritta2017}
Martina Sgritta, Francesca Locatelli, Teresa Soda, Francesca Prestori, and
  Egidio~Ugo D'Angelo.
\newblock {Hebbian spike-timing dependent plasticity at the cerebellar input
  stage}.
\newblock {\em Journal of Neuroscience}, 37(11):2809--2823, mar 2017.
\newblock \href {https://doi.org/10.1523/JNEUROSCI.2079-16.2016}
  {\path{doi:10.1523/JNEUROSCI.2079-16.2016}}.

\bibitem{Abbott2000}
L.~F. Abbott and Sacha~B. Nelson.
\newblock {Synaptic plasticity: Taming the beast}.
\newblock {\em Nature Neuroscience}, 3(11s):1178--1183, 2000.
\newblock \href {https://doi.org/10.1038/81453} {\path{doi:10.1038/81453}}.

\bibitem{Mishra2016}
Rajiv~K Mishra, Sooyun Kim, Segundo~J Guzman, and Peter Jonas.
\newblock {Symmetric spike timing-dependent plasticity at CA3-CA3 synapses
  optimizes storage and recall in autoassociative networks}.
\newblock {\em Nature Communications}, 7, 2016.
\newblock \href {https://doi.org/10.1038/ncomms11552}
  {\path{doi:10.1038/ncomms11552}}.

\bibitem{Turrigiano2000}
Gina~G Turrigiano and Sacha~B Nelson.
\newblock Hebb and homeostasis in neuronal plasticity.
\newblock {\em Current Opinion in Neurobiology}, 10(3):358--364, 2000.
\newblock URL:
  \url{https://www.sciencedirect.com/science/article/pii/S095943880000091X},
  \href {https://doi.org/10.1016/S0959-4388(00)00091-X}
  {\path{doi:10.1016/S0959-4388(00)00091-X}}.

\bibitem{Kricheldorff2022}
Julius Kricheldorff, Katharina Göke, Maximilian Kiebs, Florian~H. Kasten,
  Christoph~S. Herrmann, Karsten Witt, and Rene Hurlemann.
\newblock Evidence of neuroplastic changes after transcranial magnetic,
  electric, and deep brain stimulation.
\newblock {\em Brain Sciences}, 12(7), 2022.
\newblock URL: \url{https://www.mdpi.com/2076-3425/12/7/929}, \href
  {https://doi.org/10.3390/brainsci12070929}
  {\path{doi:10.3390/brainsci12070929}}.

\bibitem{Starosta2022}
Michał Starosta, Natalia Cichoń, Joanna Saluk-Bijak, and Elżbieta Miller.
\newblock Benefits from repetitive transcranial magnetic stimulation in
  post-stroke rehabilitation.
\newblock {\em Journal of Clinical Medicine}, 11(8), 2022.
\newblock URL: \url{https://www.mdpi.com/2077-0383/11/8/2149}, \href
  {https://doi.org/10.3390/jcm11082149} {\path{doi:10.3390/jcm11082149}}.

\bibitem{Fitzgerald2021}
Paul~B. Fitzgerald.
\newblock Targeting repetitive transcranial magnetic stimulation in depression:
  do we really know what we are stimulating and how best to do it?
\newblock {\em Brain Stimulation}, 14(3):730--736, 2021.
\newblock URL:
  \url{https://www.sciencedirect.com/science/article/pii/S1935861X21000887},
  \href {https://doi.org/10.1016/j.brs.2021.04.018}
  {\path{doi:10.1016/j.brs.2021.04.018}}.

\bibitem{Hebb1949}
Donald~O. Hebb.
\newblock {\em The organization of behavior; a neuropsychological theory}.
\newblock Wiley, 1949.

\bibitem{Taherkhani2020}
Aboozar Taherkhani, Ammar Belatreche, Yuhua Li, Georgina Cosma, Liam~P.
  Maguire, and T.M. McGinnity.
\newblock A review of learning in biologically plausible spiking neural
  networks.
\newblock {\em Neural Networks}, 122:253--272, 2020.
\newblock URL:
  \url{https://www.sciencedirect.com/science/article/pii/S0893608019303181},
  \href {https://doi.org/10.1016/j.neunet.2019.09.036}
  {\path{doi:10.1016/j.neunet.2019.09.036}}.

\bibitem{berner2023adaptive}
Rico Berner, Thilo Gross, Christian Kuehn, J{\"u}rgen Kurths, and Serhiy
  Yanchuk.
\newblock Adaptive dynamical networks.
\newblock {\em Physics Reports}, 1031:1--59, 2023.

\bibitem{Aoki2009}
Takaaki Aoki and Toshio Aoyagi.
\newblock {Co-evolution of phases and connection strengths in a network of
  phase oscillators}.
\newblock {\em Physical Review Letters}, 102(3), 2009.
\newblock \href {https://doi.org/10.1103/PhysRevLett.102.034101}
  {\path{doi:10.1103/PhysRevLett.102.034101}}.

\bibitem{aoki2015self}
Takaaki Aoki.
\newblock Self-organization of a recurrent network under ongoing synaptic
  plasticity.
\newblock {\em Neural Networks}, 62:11--19, 2015.

\bibitem{skardal2014complex}
Per~Sebastian Skardal, Dane Taylor, and Juan~G Restrepo.
\newblock Complex macroscopic behavior in systems of phase oscillators with
  adaptive coupling.
\newblock {\em Physica D: Nonlinear Phenomena}, 267:27--35, 2014.

\bibitem{Duchet2023}
Benoit Duchet, Christian Bick, and Áine Byrne.
\newblock {Mean-Field Approximations With Adaptive Coupling for Networks With
  Spike-Timing-Dependent Plasticity}.
\newblock {\em Neural Computation}, 35(9):1481--1528, 08 2023.
\newblock URL: \url{https://doi.org/10.1162/neco\_a\_01601}, \href
  {https://arxiv.org/abs/https://direct.mit.edu/neco/article-pdf/35/9/1481/2152773/neco\_a\_01601.pdf}
  {\path{arXiv:https://direct.mit.edu/neco/article-pdf/35/9/1481/2152773/neco\_a\_01601.pdf}},
  \href {https://doi.org/10.1162/neco_a_01601}
  {\path{doi:10.1162/neco_a_01601}}.

\bibitem{Ott2008}
Edward Ott and Thomas~M Antonsen.
\newblock {Low dimensional behavior of large systems of globally coupled
  oscillators}.
\newblock {\em Chaos: An Interdisciplinary Journal of Nonlinear Science},
  18(3):37113, 2008.
\newblock \href {https://doi.org/10.1063/1.2930766}
  {\path{doi:10.1063/1.2930766}}.

\bibitem{Luke2013}
Tanushree~B. Luke, Ernest Barreto, and Paul So.
\newblock {Complete Classification of the Macroscopic Behavior of a
  Heterogeneous Network of Theta Neurons}.
\newblock {\em Neural Computation}, 25(12):3207--3234, 12 2013.
\newblock URL: \url{https://doi.org/10.1162/NECO\_a\_00525}, \href
  {https://arxiv.org/abs/https://direct.mit.edu/neco/article-pdf/25/12/3207/905073/neco\_a\_00525.pdf}
  {\path{arXiv:https://direct.mit.edu/neco/article-pdf/25/12/3207/905073/neco\_a\_00525.pdf}},
  \href {https://doi.org/10.1162/NECO_a_00525}
  {\path{doi:10.1162/NECO_a_00525}}.

\bibitem{Byrne2017}
{\'A}ine Byrne, Matthew~J. Brookes, and Stephen Coombes.
\newblock A mean field model for movement induced changes in the beta rhythm.
\newblock {\em Journal of Computational Neuroscience}, 43(2):143--158, 2017.
\newblock \href {https://doi.org/10.1007/s10827-017-0655-7}
  {\path{doi:10.1007/s10827-017-0655-7}}.

\bibitem{Coombes2019}
Stephen Coombes and {\'{A}}ine Byrne.
\newblock {\em {Next Generation Neural Mass Models}}, pages 1--16.
\newblock Springer International Publishing, Cham, 2019.
\newblock \href {https://doi.org/10.1007/978-3-319-71048-8\_1}
  {\path{doi:10.1007/978-3-319-71048-8\_1}}.

\bibitem{Byrne2020}
{\'{A}}ine Byrne, Reuben~D. O'Dea, Michael Forrester, James Ross, and Stephen
  Coombes.
\newblock {Next-generation neural mass and field modeling}.
\newblock {\em Journal of neurophysiology}, 123(2):726--742, 2020.
\newblock URL: \url{www.jn.org}, \href {https://doi.org/10.1152/jn.00406.2019}
  {\path{doi:10.1152/jn.00406.2019}}.

\bibitem{Montbrio2015}
Ernest Montbri\'o, Diego Paz\'o, and Alex Roxin.
\newblock Macroscopic description for networks of spiking neurons.
\newblock {\em Phys. Rev. X}, 5:021028, Jun 2015.
\newblock \href {https://doi.org/10.1103/PhysRevX.5.021028}
  {\path{doi:10.1103/PhysRevX.5.021028}}.

\bibitem{Laing2014}
Carlo~R. Laing.
\newblock Derivation of a neural field model from a network of theta neurons.
\newblock {\em Phys. Rev. E}, 90:010901, Jul 2014.
\newblock URL: \url{https://link.aps.org/doi/10.1103/PhysRevE.90.010901}, \href
  {https://doi.org/10.1103/PhysRevE.90.010901}
  {\path{doi:10.1103/PhysRevE.90.010901}}.

\bibitem{Laing2015}
Carlo~R. Laing.
\newblock Exact neural fields incorporating gap junctions.
\newblock {\em SIAM Journal on Applied Dynamical Systems}, 14(4):1899--1929,
  2015.
\newblock \href {https://arxiv.org/abs/https://doi.org/10.1137/15M1011287}
  {\path{arXiv:https://doi.org/10.1137/15M1011287}}, \href
  {https://doi.org/10.1137/15M1011287} {\path{doi:10.1137/15M1011287}}.

\bibitem{Taher2020}
Halgurd Taher, Alessandro Torcini, and Simona Olmi.
\newblock {Exact neural mass model for synaptic-based working memory}.
\newblock {\em PLOS Computational Biology}, 16(12):e1008533, dec 2020.
\newblock \href {https://doi.org/10.1371/journal.pcbi.1008533}
  {\path{doi:10.1371/journal.pcbi.1008533}}.

\bibitem{Gast2020}
Richard Gast, Helmut Schmidt, and Thomas~R. Kn{\"{o}}sche.
\newblock {A mean-field description of bursting dynamics in spiking neural
  networks with short-term adaptation}.
\newblock {\em Neural Computation}, 32(9):1615--1634, 2020.
\newblock \href {https://doi.org/10.1162/neco\_a\_01300}
  {\path{doi:10.1162/neco\_a\_01300}}.

\bibitem{Gast2021a}
Richard Gast, Thomas~R Kn{\"{o}}sche, and Helmut Schmidt.
\newblock {Mean-field approximations of networks of spiking neurons with
  short-term synaptic plasticity}.
\newblock {\em Physical Review E}, 104:44310, 2021.
\newblock \href {https://doi.org/10.1103/physreve.104.044310}
  {\path{doi:10.1103/physreve.104.044310}}.

\bibitem{Seliger2002}
Philip Seliger, Stephen~C Young, and Lev~S Tsimring.
\newblock {Plasticity and learning in a network of coupled phase oscillators}.
\newblock {\em Physical Review E - Statistical Physics, Plasmas, Fluids, and
  Related Interdisciplinary Topics}, 65(4):7, 2002.
\newblock \href {https://doi.org/10.1103/PhysRevE.65.041906}
  {\path{doi:10.1103/PhysRevE.65.041906}}.

\bibitem{dhooge2008new}
Annick Dhooge, Willy Govaerts, Yu~A Kuznetsov, H~Ga{\'e}tan~Ellart Meijer, and
  Bart Sautois.
\newblock New features of the software matcont for bifurcation analysis of
  dynamical systems.
\newblock {\em Mathematical and Computer Modelling of Dynamical Systems},
  14(2):147--175, 2008.

\bibitem{arnold2012geometrical}
Vladimir~Igorevich Arnold.
\newblock {\em Geometrical methods in the theory of ordinary differential
  equations}, volume 250.
\newblock Springer Science \& Business Media, 2012.

\bibitem{kuznetsov1998elements}
Yuri~A Kuznetsov, Iu~A Kuznetsov, and Y~Kuznetsov.
\newblock {\em Elements of applied bifurcation theory}, volume 112.
\newblock Springer, 1998.

\bibitem{cumming1987deviations}
Andrew Cumming and Paul~S Linsay.
\newblock Deviations from universality in the transition from quasiperiodicity
  to chaos.
\newblock {\em Physical review letters}, 59(15):1633, 1987.

\bibitem{keane2016investigating}
Andrew Keane, Bernd Krauskopf, and Claire Postlethwaite.
\newblock Investigating irregular behavior in a model for the {E}l {N}i{\~n}o
  {S}outhern {O}scillation with positive and negative delayed feedback.
\newblock {\em SIAM Journal on Applied Dynamical Systems}, 15(3):1656--1689,
  2016.

\bibitem{heltberg2021tale}
Mathias~L Heltberg, Sandeep Krishna, Leo~P Kadanoff, and Mogens~H Jensen.
\newblock A tale of two rhythms: Locked clocks and chaos in biology.
\newblock {\em Cell Systems}, 12(4):291--303, 2021.

\bibitem{grebogi1983crises}
Celso Grebogi, Edward Ott, and James~A Yorke.
\newblock Crises, sudden changes in chaotic attractors, and transient chaos.
\newblock {\em Physica D: Nonlinear Phenomena}, 7(1-3):181--200, 1983.

\bibitem{Lucken2016}
Leonhard L{\"{u}}cken, Oleksandr~V Popovych, Peter~A Tass, and Serhiy Yanchuk.
\newblock {Noise-enhanced coupling between two oscillators with long-term
  plasticity}.
\newblock {\em Physical Review E}, 93(3), 2016.
\newblock \href {https://doi.org/10.1103/PhysRevE.93.032210}
  {\path{doi:10.1103/PhysRevE.93.032210}}.

\bibitem{Berner2019}
Rico Berner, Eckehard Sch{\"{o}}ll, and Serhiy Yanchuk.
\newblock {Multiclusters in networks of adaptively coupled phase oscillators}.
\newblock {\em SIAM Journal on Applied Dynamical Systems}, 18(4):2227--2266,
  2019.
\newblock \href {https://doi.org/10.1137/18M1210150}
  {\path{doi:10.1137/18M1210150}}.

\bibitem{Citri2008}
Ami Citri and Robert~C Malenka.
\newblock Synaptic plasticity: Multiple forms, functions, and mechanisms.
\newblock {\em Neuropsychopharmacology}, 33(1):18--41, 2008.
\newblock \href {https://doi.org/10.1038/sj.npp.1301559}
  {\path{doi:10.1038/sj.npp.1301559}}.

\bibitem{Turrigiano2012}
Gina Turrigiano.
\newblock {Homeostatic synaptic plasticity: local and global mechanisms for
  stabilizing neuronal function}.
\newblock {\em Cold Spring Harbr Perspectives in Biology}, 4, 2012.
\newblock \href {https://doi.org/10.1101/cshperspect.a005736}
  {\path{doi:10.1101/cshperspect.a005736}}.

\bibitem{Berberich2017}
Sven Berberich, J{\"o}rg Pohle, Marie Pollard, Janet Barroso-Flores, and Georg
  K{\"o}hr.
\newblock Interplay between global and pathway-specific synaptic plasticity in
  ca1 pyramidal cells.
\newblock {\em Scientific Reports}, 7(1):17040, 2017.
\newblock \href {https://doi.org/10.1038/s41598-017-17161-z}
  {\path{doi:10.1038/s41598-017-17161-z}}.

\bibitem{keane2018chenciner}
Andrew Keane and Bernd Krauskopf.
\newblock Chenciner bubbles and torus break-up in a periodically forced delay
  differential equation.
\newblock {\em Nonlinearity}, 31(6):R165, 2018.

\bibitem{hasan2018saddle}
Cris~R Hasan, Bernd Krauskopf, and Hinke~M Osinga.
\newblock Saddle slow manifolds and canard orbits in $\mathbb{R}^4$ and
  application to the full hodgkin--huxley model.
\newblock {\em The Journal of Mathematical Neuroscience}, 8:1--48, 2018.

\bibitem{dankowicz2020multidimensional}
Harry Dankowicz, Yuqing Wang, Frank Schilder, and Michael~E Henderson.
\newblock Multidimensional manifold continuation for adaptive boundary-value
  problems.
\newblock {\em Journal of Computational and Nonlinear Dynamics}, 15(5):051002,
  2020.

\bibitem{fenichel1979geometric}
Neil Fenichel.
\newblock Geometric singular perturbation theory for ordinary differential
  equations.
\newblock {\em Journal of differential equations}, 31(1):53--98, 1979.

\bibitem{Jones1995}
Christopher K. R.~T. Jones.
\newblock {\em Geometric singular perturbation theory}, pages 44--118.
\newblock Springer Berlin Heidelberg, Berlin, Heidelberg, 1995.
\newblock \href {https://doi.org/10.1007/BFb0095239}
  {\path{doi:10.1007/BFb0095239}}.

\end{thebibliography}

\appendix

\section{Quadratic integrate-and-fire model}\label{app:QIF}
We consider a network of quadratic integrate-and-fire (QIF) neurons 
\begin{align}
\tau_m \FD{v_j}{t} = v_j^2 + \eta_j + s_j(v_{\rm syn}-v_j).
\end{align}
Using the transformation $v_j=\tan \theta_j /2$, we derive the equivalent phase dynamics \eqref{eq:theta_network}.

\end{document}